\documentclass[%
 reprint,
superscriptaddress,
longbibliography,
 amsmath,amssymb,
 aps,
prb,
]{revtex4-2}
\usepackage[english]{babel}

\usepackage{graphicx}
\graphicspath{ {./figures/} }
\usepackage{amsmath, amssymb}
\usepackage{dcolumn}
\usepackage{bm}
\usepackage{physics}
\usepackage[colorlinks=true,citecolor=blue,linkcolor=red,linktocpage=true]{hyperref}
\usepackage{xcolor}

\definecolor{AN}{rgb}{0.5,0.1,0.3}

\begin{document}
\preprint{APS/123-QED}

\title{Topological edge flows drive macroscopic re-organization in magnetic colloids}
\author{Aleksandra~Nelson}
\thanks{These authors contributed equally to this work}
\affiliation{Center for Theoretical Biological Physics, Rice University, Houston TX, USA}
\author{Dana~M.~Lobmeyer}
\thanks{These authors contributed equally to this work}
\affiliation{Department of Chemical and Biomolecular Engineering, Rice University, Houston TX, USA}
\author{Sibani~L.~Biswal}
\affiliation{Department of Chemical and Biomolecular Engineering, Rice University, Houston
TX, USA}
\author{Evelyn~Tang}
\affiliation{Center for Theoretical Biological Physics, Rice University, Houston TX, USA}
\affiliation{%
 Department of Physics and Astronomy, Rice University, Houston TX, USA
}
\date{\today}

\begin{abstract}
Magnetic colloids can be driven with time-varying fields to form clusters and voids that re-organize over vastly different timescales. However, the driving force behind these non-equilibrium dynamics is not well-understood. Here, we introduce a topological framework that predicts protected edge flows despite strong thermal motion. Notably, these edge flows produce shear stress that creates global rotation of clusters but not of voids. We verify this theory experimentally using micron-sized super-paramagnetic colloids to demonstrate {these emergent physical predictions and show how they drive system re-organization differentially at long timescales.} Our results elucidate fundamental principles that shape and control non-equilibrium colloidal aggregates.
\end{abstract}

\maketitle

\section{Introduction}
The ability to tune particle-particle interactions at the colloidal scale in-situ provides a promising platform for engineering materials with uniquely responsive properties \cite{solomon2018, li2022}. Such responsiveness has been accomplished primarily through temperature \cite{liao2018} and/or external fields \cite{komarov2020, alharraq2022}. Magnetic external fields enable a range of fine-tuned interactions \cite{osterman2009, soheilian2018}, making them valuable in both biological and non-biological applications due to their safety and simplicity.  These applications range from experimental models for atomic and molecular systems to the formation of transport vehicles for enhanced drug delivery \cite{joshi2022, mattich2023, bishop2023}. Magnetic colloids often display cluster rotation that has been harnessed for drug delivery applications \cite{tasci2017, zimmermann2022}, and individual particle rotation has been utilized for various micro-scale transport \cite{martinez-pedrero2015, massana-cid2019}. Furthermore, they can produce chiral fluids with unique properties such as odd viscosity \cite{soni2019}. 

While the behavior of individual particles under rotating magnetic fields has been well studied \cite{du2013, cunha2024}, there is a lack of theory to describe the emergence of cluster rotation. Some works have attributed this rotation to magnetic cluster anisotropy \cite{tierno2007}, or to the hydrodynamic effects between individual rotating particles \cite{jager2013}. However, the equilibrium descriptions often used to describe collective behavior in these systems \cite{hilou2018} are insufficient for such non-equilibrium dynamics. While there have been numerical studies detailing order from particle interactions \cite{shen2023, caporusso2024}, it would be advantageous to have a theoretical framework that can be applied across various material properties and platforms. 

Topological theory has become a recent theoretical framework of interest because it predicts macroscopic emergent behavior from microscopic interactions across various disparate platforms. Such platforms include electronic systems \cite{hasan2010, moore2010, qi2011}, photonic crystals \cite{lu2014, ozawa2019}, mechanical lattices \cite{huber2016, mao2018, zheng2022} and stochastic networks \cite{murugan2017, dasbiswas2018, tang2021, zheng2024}. Furthermore, topological theory predicts emergent edge states and currents that are robust to material defects and deformations \cite{hasan2010, moore2010, qi2011}. In the realm of soft and active matter \cite{shankar2022, serra2020, mecke2024}, these can take the form of protected edge flows insensitive to defects or disorder in the material. Striking examples include fluid and particle flows on the interfaces of active metamaterials \cite{shankar2017, dasbiswas2018, sone2019, souslov2019, sone2020}, equatorial waves in the ocean \cite{delplace2017},  and chiral edge currents in nematic cell monolayers \cite{yashunsky2022}. 

In this work, we introduce a topological hydrodynamic theory to account for edge flows in colloidal assemblies under a rotating magnetic field and examine our theory predictions in an equivalent experimental setup. While this approach was initially developed for chiral particles at constant density within fixed walls \cite{dasbiswas2018, yang2020}, we adapt this framework for experimentally relevant Brownian super-paramagnetic colloids that self-assemble into regions of high and low density. We demonstrate that topological edge flows are insensitive to shape changes, presenting not only in isolated clusters but also along novel empty void spaces in dense particle sheets. {Using} the developed theory we {provide} predictions of macroscopic features of the system, including edge 
{speeds} and decay length. {Notably, we show how the edge flows lead to shear stress that causes the rotation of clusters but not of voids, which drives system re-organization differentially on long timescales relevant for domain coarsening.
}

\section{Magnetic colloidal assemblies contain prominent edge flows}

\begin{figure}
    \centering    \includegraphics{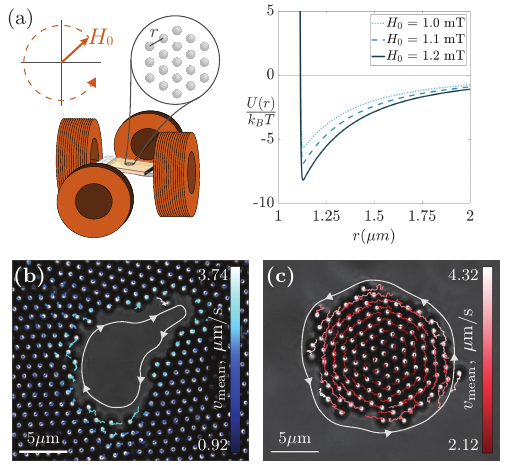}
    \caption{\textbf{Magnetically driven colloids form voids and clusters with edge flows.} (a) Left: Experimental setup with superparamagnetic colloids at the center of a rotating magnetic field (orange arrow), created by two pairs of magnetic coils. Right: The interaction potential between two particles as a function of interparticle distance, $r$, for three magnetic field strengths. (b, c) Colloid trajectories over 30 frames mapped onto microscopy images of (b) a crystalline sheet containing an empty void and (c) a cluster at $H_0=1.2$ mT. The trajectories are colored by the magnitude of the mean particle {speed}, $v_\mathrm{mean}$. The direction of edge flows is indicated with white arrows.}
    \label{fig:1}
\end{figure}

{Magnetically assembled sheets and clusters are created by subjecting a suspension of 1 $\mu$m super-paramagnetic particles (Dynabeads MyOne Carboxylic Acid, Invitrogen) in 10mM NaCl to a continuously rotating magnetic field. The suspension is held within an epoxy sealed chamber made of negatively charged glass slides and a Parafilm spacer. Since the particles are denser than the fluid and also negatively charged, they settle to a quasi-two-dimensional plane just above the substrate. The magnetic field is created by running a sine and cosine wave through two pairs of magnetic coils set orthogonal to one another, see Fig.~\ref{fig:1}(a). Upon application of a 20Hz field, the particles acquire magnetic dipoles and begin to interact according to the time-averaged interaction potential for super-paramagnetic particles in a continuously rotating magnetic field, that was theoretically derived and experimentally measured in \cite{du2013}. This two-dimensional interaction potential, shown on the right of Fig.~\ref{fig:1}(a) for different field strengths, is composed of an electrostatic short-range repulsion and a long-range magnetic attraction. The short-range repulsion can be tuned by the salt solution, while the long-range attraction is tuned by the magnetic field strength, $H_0$. Under this relatively fast frequency field, particle assemblies with well defined, tunable order are possible \cite{spatafora-salazar2021}.}

Sheets {of particles} containing {empty} voids or {individual particle} clusters in empty space are achieved by varying the particle concentration under the same magnetic field conditions. {This ability to tune morphology is a direct result of phase separation and energy minimization, where the system separates into particle rich and particle poor regions and seeks to minimize energy by reducing the interface length \cite{hilou2020}.} At a relatively high particle concentration of 3-4mg/ml {and corresponding particle surface coverage of 30-40\%}, sheets of particles form as a continuous polycrystalline phase containing empty voids (Fig.~\ref{fig:1}(b)). In contrast, a reduced concentration of 1-1.25mg/ml {(10-12\% particle surface coverage)} achieves isolated crystalline clusters (Fig.~\ref{fig:1}(c)). In the cluster phase, single crystalline domains {(clusters)} exist {physically} separate from one another and are free to move. In the void phase, however, many crystalline domains exist within the constrained bulk, separated by disordered boundaries \cite{lobmeyer2022}. The relative movement of these constrained domains in the void phase is much reduced compared to the cluster phase.

At the edge of voids and on the perimeter of clusters, we observe that particles exhibit
edge flows, as shown by the particle trajectories in Figs.~\ref{fig:1}(b) and (c). Edge flows are composed of the particles along the edge which translate faster than particles in the bulk, as reflected in the mean {speed}, $v_\textnormal{mean}$, computed in the IMARIS software as the total 
{path}
traveled over the length of time tracked. With our frame rate of 30fps and density of about 1 particle$/\mu \mathrm{m}$, several frames are captured during each translation of a particle length for our observed speeds of a few  $\mu \mathrm{m}/\mathrm{s}$.

In addition, all particles have a constant background thermal speed computed as an average path traveled due to the Brownian motion. Brownian particles display structure on all length scales, so while the total path of a Brownian trajectory depends on the sampling rate, this rate will affect the magnitude of the plateau but not the existence of the plateau itself. We estimate our noise plateaus to be about 2 $\mu \mathrm{m}/\mathrm{s}$, which is adequately captured by our sampling rate as discussed above. Notably, most of the data lie in the exponentially-decaying portion, from which we can see that the faster translation at the edge occurs regardless of magnetic field strength. Along voids, all edge particles translate but only a few particles move significantly at any one point in time. Meanwhile, the polycrystalline bulk remains fairly stationary, beyond thermal motion, over the course of an experiment (approximately 1 minute). In contrast, all particles along the edge of clusters move continuously and the entire cluster rotates as a result.

\section{Topological origin of edge flows}

\begin{figure*}
    \centering
    \includegraphics{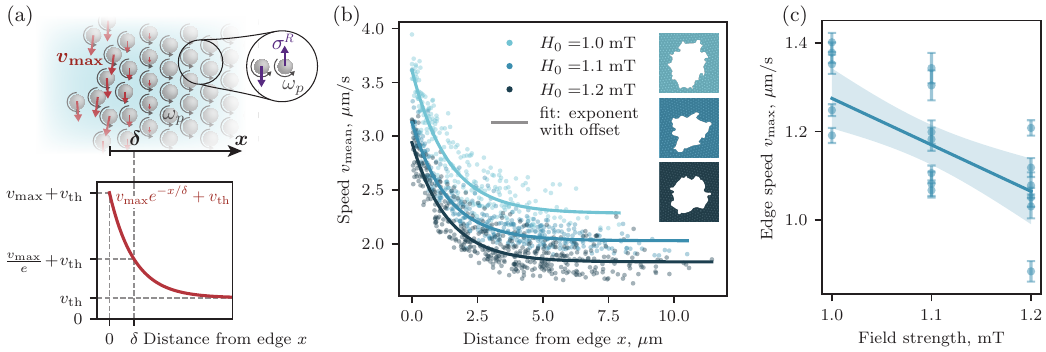}
    \caption{\textbf{Topological theory predicts edge flows in voids.} (a) Top: A graphic of angular velocity, $\omega_p$ (gray arrows), coupling to linear velocity, $v$, through rotational stress, $\sigma_R$ (purple arrows), that creates unbalanced stress at the edge, inducing a topologically protected flow of particles (red arrows). Bottom: The predicted velocity profile from Eq.~\ref{eq:model} with decay length, $\delta$, and thermal velocity, $v_\mathrm{th}$.
    (b) The observed mean 
    {speed}, $v_\mathrm{mean}$, of particles around three voids plotted against their distance from the edge for three magnetic field strengths. The solid lines indicate fits to Eq.~\eqref{eq:model} and the right insets depict each void's shape. (c) The maximal edge 
    {speeds}, $v_\mathrm{max}$, for eighteen different voids across the three field strengths. The solid line indicates a linear fit of the data with a 95\% confidence interval (shaded).
    }
    \label{fig:2}
\end{figure*}

To develop our theoretical framework, we begin from a hydrodynamic description of colloidal assemblies treated as continuous media \cite{tsai2005, vanzuiden2016, soni2019}.
Under a rotating magnetic field, magnetic torque imparts particles with an angular velocity, $\omega_p$, in the direction of field rotation \cite{cunha2024}.
Hydrodynamic interactions couple this angular velocity to a linear velocity, $\bm{v}$, through the rotational stress tensor, $\sigma_{ij}^R=\eta_R\epsilon_{ij}(2\omega_p-\zeta)$, where $\zeta=(\nabla\times\bm{v})_z$ is vorticity that describes circulating flows, $\eta_R$ is rotational viscosity, and $\epsilon_{ij}$ is the antisymmetric Levi-Civita tensor \cite{tsai2005}. 
In the colloidal bulk, rotational stress from all neighboring particles cancels out, such that vorticity vanishes (Fig.~\ref{fig:2}(a)). However, for particles on the edge, the rotational stress is present only from one side. This stress imbalance induces a finite vorticity which corresponds to the linear velocity of an edge flow.

When rotational viscosity $\eta_R$ is small compared to linear viscosity $\eta$, the angular velocity and vorticity decouple, such that vorticity follows equation \ref{eq:hydrodyn_eq}, \cite{soni2019}
\begin{equation}
    (\nabla^2-\delta^{-2})\zeta=0,
    \label{eq:hydrodyn_eq}
\end{equation}
where the length scale $\delta=\sqrt{(\eta+\eta_R)/\Gamma_v}$ is determined by the two viscosities and friction with the substrate, $\Gamma_v$ (see Appendix~\ref{app:hydrodynamics} for details). Notably, this operator has the same form as the square of a well-known topological Hamiltonian, the Dirac Hamiltonian (see Appendix~\ref{app:topology} for details). Eigenvectors of this Hamiltonian have a winding number known as the Chern number, which yields edge states robust to defects and random perturbations \cite{hasan2010, bernevig2013}.

At the interface between particle-rich and particle-poor regions, the Chern number changes its value, inducing a state with zero eigenvalue localized at the interface \cite{dasbiswas2018}. This state is exactly the steady state solution of Eq.~\eqref{eq:hydrodyn_eq}, since these operators have the same zero eigenvalue solution. This solution yields a topologically-protected steady state where vorticity $\zeta(x)$ decays exponentially away from the edge, $x$. To obtain the 
{particles speed} $v(x)$ along the edge, we integrate $\zeta(x)$ (see Appendix~\ref{app:topology} for details) to obtain,
\begin{equation}
    v(x)= v_\mathrm{max} e^{-x/\delta}+v_\mathrm{th},
    \label{eq:model}
\end{equation}
where $v_\mathrm{max}$ is the maximal edge 
{speed}, $\delta$ is the decay length and $v_\mathrm{th}$ is the average thermal motion (see Fig.~\ref{fig:2}(a)).

To verify this prediction, we fit Eq. \ref{eq:model} to experimentally observed flows of particles at the edges of voids and clusters. Since voids exhibit much less global rotation compared to clusters, the edge flows can be more readily isolated. Particle {flows}
are characterized by their mean
{speed},
$v_\mathrm{mean}$, and their mean distance from the edge, $x$, where $x=0$ marks the edge. The edge is identified with the IMARIS software's surface function, which identifies the empty void space based on image intensity values. 

\begin{figure*}
    \centering
    \includegraphics{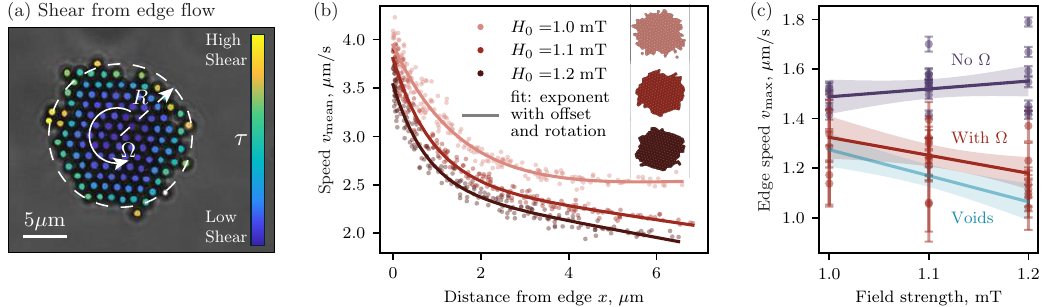}
    \caption{\textbf{Edge flows induce cluster rotation.} (a) Microscopy image of a cluster with radius $R$, rotating at an angular velocity $\Omega$. Particles are colored by the magnitude of the relative shear stress, $\tau$, averaged over 16 seconds.
    (b) The experimentally measured mean 
    {speed}, $v_\mathrm{mean}$, of particles plotted against distance from the edge for three clusters at different magnetic field strengths. The solid lines indicate fits to Eq.~\eqref{eq:rotfit} and the right insets depict each cluster's shape.  
    (c) The maximal edge 
    {speeds}, $v_\mathrm{max}$, for twenty-one unique clusters across three field strengths for fits with (red) and without (purple) bulk rotation. The solid lines indicate linear fits of the data with 95\% confidence intervals (shaded). The same fit for voids in Fig~\ref{fig:2}(c) is shown with a blue line for reference.
     }
    \label{fig:3}
\end{figure*}

Figure~\ref{fig:2}(b) shows Eq. \ref{eq:model} fit to the experimentally measured $v_\mathrm{mean}(x)$ for three voids under increasing values of magnetic field strength. At all field strengths, the average thermal motion is significant, corresponding to the value at which the edge 
{speeds}
plateau ($v_\mathrm{th}$). However, despite this significant thermal motion, we find the R-squared statistic of fits to be 0.86, 0.89, and 0.83 for the field strengths $H_0=1.0$, $1.1$, and $1.2$ mT, respectively, showing that the topological edge flows are robust to strong noise. As the magnetic field increases,  the magnetic interactions are expected to dominate over the hydrodynamic effects. This magnetic domination is confirmed by the decrease in the fitted maximal edge 
{speed} with field strength,  showcased in Fig.~\ref{fig:2}(c) for five ($1.0$mT) , seven ($1.1$mT), and six voids ($1.2$mT). Since the increased magnetic interactions promote {a stronger attraction between particles that densifies the cluster}, the effect of hydrodynamics is reduced, thereby reducing the strength of edge flows. This is consistent with the decrease in $v_\mathrm{th}$ with the magnetic field strength in Fig.~\ref{fig:2}(b), since stronger magnetic interactions also decrease random motion.

\section{Edge flows induce cluster rotation}

In contrast to the relatively stationary voids, clusters display global rotation. To understand the origin of this rotation, we examine the shear stress across a cluster. Shear stress is computed per particle, as the relative shear of one particle on its neighboring particle that is interior to the cluster,
\begin{equation}
\tau_i = \left\langle \frac{v_i - v_\mathrm{interior}}{\Delta x} \right\rangle,
\label{eq:shear}
\end{equation} 
where $v_i$ is the 
{speed}
of the particle of interest, $v_\mathrm{interior}$ is the 
{speed}
of the neighboring particle most interior, $x$ is the distance between the two particles and $\langle \rangle$ is a time average. Since the viscosity is assumed constant, it does not appear in the expression above. 

In comparing relative shear across a cluster, we find that the shear stress pattern is greatest at the edge and decreases as it moves toward a bulk center where shear stress values are the same (see Fig.~\ref{fig:3}(a)). This shear stress pattern reflects that of the edge flow pattern, suggesting that edge flows impart a shear stress at the edge of the cluster that causes rotation of a center bulk, akin to rigid-body rotation. To include rigid-body rotation in our analysis, we add a linear term to our steady-state solution in Eq. \eqref{eq:model}, in line with previous works \cite{yan2015,soni2019}:
\begin{equation}
    v(x)=v_\textrm{max}e^{-x/\delta}+v_\textnormal{th}+\Omega(R-x).
    \label{eq:rotfit}
\end{equation}
Here, $\Omega$ is the bulk cluster rotation and $R$ is the effective radius of the cluster, determined by fitting the cluster area to a representative circle. 

Equation~\ref{eq:rotfit} is fit to the experimentally measured $v_\mathrm{mean}(x)$ for three clusters in Fig.~\ref{fig:3}(b) under the same magnetic field strengths as before. As can be seen, the particle mean 
{speeds} decay exponentially away from the edge, much like in the case of the voids. However, unlike the voids, the behavior becomes linear in the bulk, reflecting the global cluster rotation. The goodness of fit is showcased in the R-squared statistic of fits being 0.96, 0.98, and 0.96 for the field strengths $H_0=1.0$, $1.1$, and $1.2$ mT, respectively.

To evaluate the importance of the rotational component in Eq.~\ref{eq:rotfit}, we compare the fit  with (Eq.~\ref{eq:rotfit}) and without a rotational term (Eq.~\eqref{eq:model}) by looking at the fitted maximal edge 
{speed} with field strength. As shown in Fig.~\ref{fig:3}(c), five ($1.0$mT), nine ($1.1$mT)), and seven ($1.2$mT)) clusters are analyzed for each case.  When the fit with the rotational term is considered, the maximal edge 
{speeds} of the clusters decrease with field strength (red line in Fig.~\ref{fig:3}(c)), in line with the trend of the voids. In contrast, when the rotational term is not considered, the opposite trend occurs and the maximal edge 
{speed} increases with field strength (purple line in Fig.~\ref{fig:3}(c)). Since the balance of hydrodynamic and magnetic interactions should remain constant with magnetic field, independent of geometry, we find the rotational term is necessary to properly describe edge flows in the cluster geometry. This indicates significant shear induced rotation of clusters in contrast to voids that are fixed in place and cannot rotate.

\begin{figure*}
    \centering
    \includegraphics{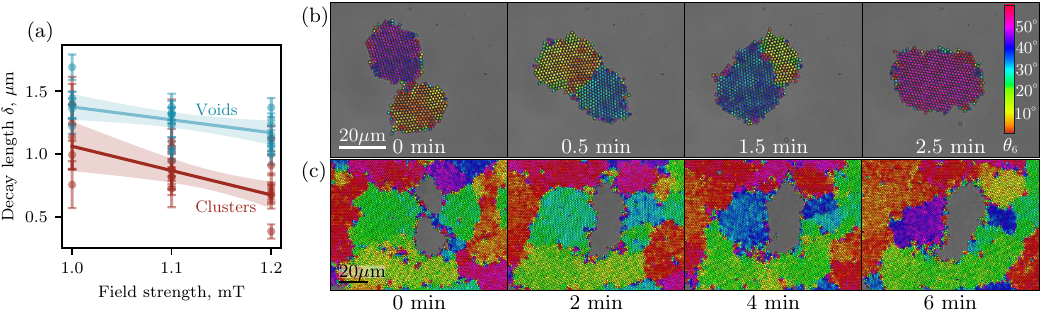}
    \caption{\textbf{Edge flows produce fast re-organization in clusters compared to voids.} (a) The decay lengths obtained from 
    {speed} fits using Eq.~\ref{eq:model} and \ref{eq:rotfit} for voids (blue dots) and clusters (red dots), respectively. The solid lines indicate linear fits with 95\% confidence intervals (shaded) (b) Microscopy images of coalescing clusters and (c) voids across four time points with an overlaid colormap of the local lattice orientation, $\theta_6$.}
    \label{fig:4}
\end{figure*}

\section{Edge flows impact  reorganization timescales}

While clusters and voids both contain edge flows, they impact global dynamics in distinctly different ways, i.e. inducing rigid-body rotation in clusters and dissipation into the bulk in voids. It follows then that properties of the edge flows should differ between the two geometries. To examine this hypothesis, we compare the decay length, $\delta$, for voids and clusters, obtained from the 
{speed} fits using Eq.~\eqref{eq:model} and Eq.~\eqref{eq:rotfit}, respectively. We find that the decay length in voids is higher than the decay length in clusters for all field strengths (Fig.~\ref{fig:4}(a)). This longer decay length in voids can be explained by system constraints. While the shear stress around clusters can be converted into cluster rotation, the restricted bulk near voids forces the shear stress at the edge to dissipate further into the bulk, resulting in a longer decay length. This system-level constraint also explains why clusters have faster moving particles at their edge, as compared to voids, as their motion comprises both edge flows and bulk rotation (Fig.~\ref{fig:3}(c)).

These differences in macroscopic properties, i.e. decay length and average 
{speed}, carry profound consequences on the re-organization of domains in clusters as compared to voids.  Using a simple dimensional analysis where time goes as length over 
{speed}, we can predict that the timescales associated with non-equilibrium behavior will be longer in voids as compared to clusters. We compare this prediction to experiments of colloidal reorganization in each geometry. 

Figures~\ref{fig:4}(b) and (c) show the reorganization of grain boundaries over time, using an overlaid colormap of local lattice orientation $\theta_6$ measured against the positive $x$-axis. In the case of the cluster geometry, two clusters coalesce to form a larger cluster with a grain boundary (Fig.~\ref{fig:4}(b)). Such disordered boundaries between  ordered regions of different crystalline orientation have been shown to translate out of clusters and form a single crystalline domain \cite{hilou2018}. In Fig.~\ref{fig:4}(b), the grain boundary translates out of the cluster within 2.5 minutes, suggesting that reorganization in clusters happens on the time scale of seconds to minutes. In contrast, grain boundaries that appear in the bulk around voids are not significantly reduced  after more than double the time (Fig.~\ref{fig:4}(c)). In fact, voids have displayed a shear-induced reorganization of grain boundaries in the bulk that occurs on the timescale of minutes to hours \cite{lobmeyer2022}.

\section{Conclusion}
In this work, we quantitatively analyze stable edge flows in noisy assemblies of super-paramagnetic particles and demonstrate that they obey the same functional form predicted by topology across two distinct geometries. In both sheets with voids and clusters, we show that edge flows create a shear stress that must dissipate into the bulk for voids, but results in rigid-body rotation in clusters. These differences produce disparate decay lengths across voids and clusters, creating contrasting timescales for their collective motion and re-organization.  Through a theoretical framework that has broad applicability \cite{shen2023, caporusso2024,yan2015}, we report how the macroscopic properties of colloidal assemblies can be controlled with magnetic field strength and geometric confinement. Ultimately, our results extend the framework of topology to noisy and self-organizing colloids \cite{soni2019,lobmeyer2022}, towards the engineering and design \cite{tjhung2018, nejad2023} of domains in configurable materials \cite{solomon2018, li2022,liao2018,osterman2009}. Our formalism opens new directions for other platforms of recent interest, such as clusters of cells that show rotational motion \cite{ascione2023, tanner2012, wang2022, cetera2014, brandstatter2023, li2024}, paving the way for deeper understanding and control of robust dynamics in adaptive soft and living matter \cite{cheng2006}.

\section{Acknowledgments}  We thank Dr. Alloysius Budi Utama of the Rice Shared Equipment Authority (SEA) for IMARIS guidance. This work was supported by the NSF Directorate for Technology, Innovation, and Partnerships (Grant No. PFI-2141112), the NSF Center for Theoretical Biological Physics (PHY-2019745), and the NSF CAREER Award (DMR-2238667). 

\appendix

\section{Hydrodynamic theory of colloidal assemblies}

\label{app:hydrodynamics}

Following Refs.~\cite{tsai2005, soni2019}, we model the  assemblies of colloidal particles using continuous hydrodynamic theory. For this we introduce two fields:
\begin{itemize}
\item \emph{angular velocity} $\omega_p(\boldsymbol{r}, t)$ of individual particles,
\item \emph{linear velocity} $\boldsymbol{v}(\boldsymbol{r}, t)$, whose curl forms \emph{vorticity} of particles $\zeta(\boldsymbol{r}, t)=(\nabla\times\boldsymbol{v})_z$.
\end{itemize} 
The dynamics of the colloidal particles is described by the momentum and angular momentum conservation laws:
\begin{align}
\rho \partial_t v_i &= -\partial_i p + \eta\nabla^2v_i + \eta_R\epsilon_{ij}\partial_j (2\omega_p-\zeta) - \Gamma_v v_i, \label{appeq:dynamic_velocity} \\
\mathcal{I}\partial_t\omega_p &= D_\omega\nabla^2\omega_p - 2\eta_R(2\omega_p - \zeta) - \Gamma_{\omega}\omega_p + T, \label{appeq:dynamic_spinning}
\end{align} 
where $\mathcal{I}$ is the moment of inertia density and $\rho$ is the particle density. 
Terms on the right side include the pressure $p$, rotational diffusion constant of the particles $D_\omega$, linear and rotational viscosities $\eta$ and $\eta_R$, linear and rotational substrate friction coefficients $\Gamma_v$ and $\Gamma_\omega$, an antisymmetric Levi-Civita symbol $\epsilon_{ij}$, and an active torque $T$ induced by the rotating magnetic field. The rotational stress tensor $\sigma_{ij}^R=\eta_R\epsilon_{ij}(2\omega_p-\zeta)$ couples the two equations \cite{tsai2005}, thus allowing the active torque $T$ to cause linear movement of particles. {To understand the form of the rotational stress tensor, we note that when two particles rotate around each other with vorticity that is equal to twice their angular velocity, there is no rotational friction between them. When these two quantities deviate from each other, the rotational stress will try to bring them closer together.}

We search for a \emph{steady states solution} which solves the following equations, 
\begin{align}
\eta\nabla^2\zeta - \eta_R\nabla^2 (2\omega_p-\zeta) - \Gamma_v \zeta=0, \label{appeq:stst_vorticity} \\
D_{\omega}\nabla^2\omega_p - 2\eta_R(2\omega_p - \zeta) - \Gamma_{\omega}\omega_p + T=0, \label{appeq:stst_spinning}
\end{align} 
where Eq.~\eqref{appeq:stst_vorticity} is obtained by taking a curl of Eq.~\eqref{appeq:dynamic_velocity}.
In the \emph{bulk} of the colloidal assembly the solution is
\begin{align}
\bar{\zeta} &=0, \label{appeq:vorticity_bulk}\\
\bar{\omega}_p &=\frac{T}{\Gamma_{\omega}+4\eta_R} \label{appeq:spinning_bulk},
\end{align}
which corresponds to a homogeneous spinning of all particles due to external torque.

Near the \emph{boundary}, the angular velocity and vorticity fields may vary to fulfill the \emph{boundary conditions}. To simplify calculations, we assume that the particles spin much faster than they rotate around each other, e.g. $\omega_p\gg\zeta$. {As we will see later, this assumption is reasonable for assemblies of magnetic colloids.} Then the angular momentum remains a constant given by Eq.~\eqref{appeq:spinning_bulk}, while the vorticity is given by the equation:
\begin{equation}
    \nabla^2\zeta - \delta^{-2}\zeta=0, \quad \delta=\sqrt{\frac{\eta+\eta_R}{\Gamma_v}}.
    \label{appeq:vortivity_deq}
\end{equation}
\section{Topological theory and edge solution}

\label{app:topology}

Notably, the operator in Eq.~\eqref{appeq:vortivity_deq} has the same form as the square of a well-known topological Hamiltonian, the Dirac Hamiltonian, i.e., $(-\nabla^2+\delta^{-2})I_2=\mathcal{H}_d^2$, where $I_2$ is a rank two identity matrix \cite{dasbiswas2018}. The Dirac Hamiltonian is $\mathcal{H}_d=m\sigma_1-i\partial_x\sigma_2-i\partial_y\sigma_3$, where $\sigma_{1,2,3}$ are the Pauli matrices and $m=\pm\delta^{-1}$. The eigenvectors of $\mathcal{H}_d$ have a winding number known as the Chern number, which yields edge states robust to defects and random perturbations \cite{hasan2010, bernevig2013}.

At the interface between particle-rich and particle-poor regions, the Chern number changes its value, inducing a state with zero eigenvalue localized at the interface \cite{dasbiswas2018}. This state is exactly the steady state solution of Eq.~\eqref{appeq:vortivity_deq}, since the two operators $\mathcal{H}_d$ and $\mathcal{H}_d^2$ have the same zero eigenvalue solution. This solution yields a topologically-protected steady state where vorticity $\zeta(x)$ decays exponentially away from the edge, $x$.

\begin{figure}
    \centering
    \includegraphics[scale=0.9]{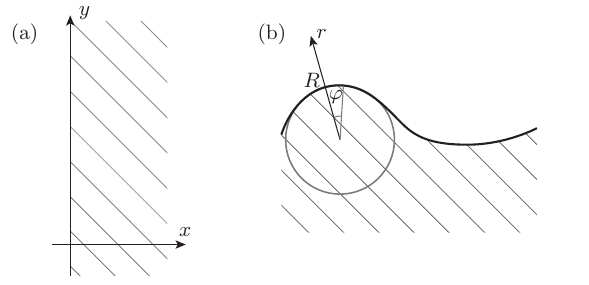}
    \caption{(a) Colloidal cluster forming a semi-infinite plane. (b) Colloidal cluster with a generic boundary. At each point of the boundary define its curvature and a polar coordinate system.}
    \label{fig:boundary_geom}
\end{figure}

To analytically find this exponentially decaying solution of Eq.~\eqref{appeq:vortivity_deq}, we adopt another simplification, namely that the boundary of colloidal assembly is close to being flat. Then we can consider the colloidal assembly as a semi-infinite plane at $x>0$ with $x=0$ line being its boundary {(see Fig.~\ref{fig:boundary_geom}(a))}. Then, the solution is $y$-independent and is
\begin{equation}
    \zeta(x)=\zeta_0 e^{-x/\delta}.
\end{equation}
We indeed observe that this solution is exponentially localized near the boundary, as predicted by the topological theory. Our initial assumption of flat boundary is justified when the radius of colloidal assemblies is much larger than the decay length of the edge flow $R\gg \delta$. {We show in the next section, that the edge flow in our clusters and voids, whose radius is a few times larger than the decay length, can be modeled with an exponential decay even if their boundary is not perfectly flat.}

Finally, the constant $\zeta_0$ can be determined by fulfilling the boundary conditions.
In the studied colloidal system, there is no hard boundary, and hence the velocity field fulfills two conditions:
\begin{enumerate}
\item the kinematic boundary condition that ensures that the boundary moves with fluid velocity,
\item vanishing normal stress.
\end{enumerate}
The second condition takes the form
\begin{align}
    \sigma_{yy}\big|_{x=0} &=-p=0, \\
    \sigma_{xy}\big|_{x=0} &= 2\omega_p\eta_R + (\eta+\eta_R)\zeta_0,
\end{align}
which holds when pressure vanishes and 
\begin{equation}
    \zeta_0=\frac{2\eta_R\omega_p}{\eta+\eta_R}.
\end{equation}
We see, that when the linear viscosity is much larger than the rotational viscosity $\eta\gg\eta_R$, the particle vorticity $\zeta$ is much smaller than the angular velocity $\omega_p$, which was our initial assumption. This relation between the viscosities is physically possible and was observed {in a similar system} in Ref.~\cite{soni2019} [cf.~Fig.~S6]. {Moreover, our colloidal particles are round in contrast to cubic particles used in Ref.~\cite{soni2019}. Therefore, we may expect the rotational viscosities to be even smaller in our case, further supporting our assumption.}

Computing the \emph{linear velocity} in this system we take into account that normal velocity {does not depend on the tangential coordinate due to translational invariance $\partial_yv_x=0$} and get
\begin{equation}
    v_y(x)=-\int\limits_x^\infty\zeta(x) dx+v_y(\infty)=\frac{2\eta_R\omega_p\delta}{\eta+\eta_R}e^{-x/\delta} + v_\mathrm{th}, \label{appeq:tangential_velocity}
\end{equation}
where the second term is the thermal motion, which can be present uniformly across the system and which equals to velocity infinitely far from the edge.
We note that the edge flow
\begin{itemize}
    \item decays exponentially with the distance from the boundary, confirming its topological interpretation;
    \item vanishes if rotational viscosity is zero $\eta_R=0$, indicating that coupling through rotational stress is necessary for edge flows.
\end{itemize}

\begin{figure}
    \centering
    \includegraphics{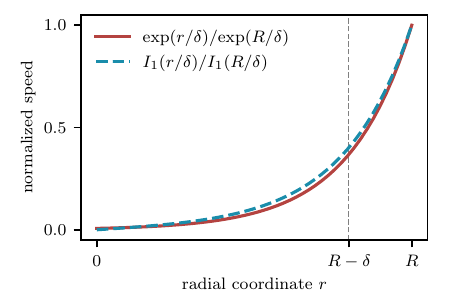}
    \caption{Edge flow profile modeled in a cluster of radius $R$ as a function of the radial coordinate $r$. Red line shows an exponential decay with decay length $\delta$. Blue dashed line shows a modified Bessel function of the first kind with decay length $\delta$. Both functions are normalized to their values at the edge of the cluster. The ratio $R/\delta=5$.}
    \label{fig:Bessel_vs_exponential}
\end{figure}
\section{Solution at curved boundary}
\label{app:curved_boundary}
\begin{figure}
    \centering
    \includegraphics[scale=0.9]{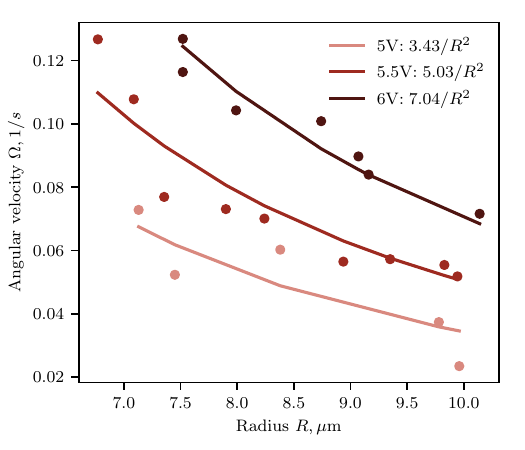}
    \caption{Plot fitted angular velocity of the cluster bulk as a function of the cluster radius for three different magnetic field strengths. Solid lines show power law fits.}
    \label{fig:bulk_rotation_vs_radius}
\end{figure}

\begin{figure*}
    \centering
    \includegraphics{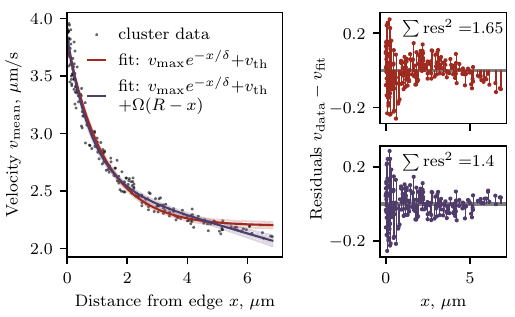}
    \caption{Left: mean {speeds} in a cluster at $H_0=1.1$ mT (black dots) are fitted by two models, without bulk rotation (red line) and with bulk rotation (purple line). Shaded areas indicate the 95\% confidence intervals. Right: residuals are plotted for the fit without bulk rotation (top panel) and with bulk rotation (bottom panel). Including bulk rotation makes residuals more symmetric around zero and the sum of their squares smaller.}
    \label{fig:cluster_two_fit}
\end{figure*}
In this Appendix, we find the edge flow solution in a more general case of a curved boundary with a curvature $\kappa=1/R$. We introduce \emph{polar coordinates} $r, \varphi$ with the center of coordinate located at the center of the circle with radius $R$ [cf.~Fig.~\ref{fig:boundary_geom}(b)]. We redefine linear velocity in these coordinates as
$v_r=\dot{r}$ and $v_\varphi=r\dot{\varphi}$ and the vorticity is expressed as $\zeta=1/r\, v_\varphi+\partial_r v_\varphi-1/r\partial_\varphi v_r$. Rewriting Eq.~\eqref{appeq:vortivity_deq} in polar coordinates we get
\begin{equation}
    r^2\partial_r^2v_\varphi+r\partial_r v_\varphi - (1+r^2/\delta^2)v_\varphi = 0,
\end{equation}
where we used that $\partial_\varphi v_r=0$ due to radial symmetry. This differential equation has a solution in the form of a \emph{modified Bessel function}
\begin{equation}
    v_\varphi(r)=\begin{cases}
    AI_1(r/\delta), &\textrm{convex boundary}\\
    BK_1(r/\delta), & \textrm{concave boundary}
    \end{cases} 
\end{equation}
where $I_1$ and $K_1$ are modified Bessel functions of the first and second kind, respectively. To find constants $A$ and $B$ we again need to solve the boundary condition
\begin{align}
    &\sigma_{rr}\big|_{r=R}=-p=\gamma/R, \\
    &\sigma_{\varphi r}\big|_{r=R}= \nonumber \\
   & = (\eta_R+\eta)\partial_r v_\varphi+(\eta_R-\eta)/r\,v_\varphi-2\eta_R\omega_p\big|_{r=R}
    =0,
\end{align}
where pressure $p$ is compensating the surface tension $\gamma$ and the tangential velocity is
\begin{equation}
    v_\varphi(r)=\begin{cases}
    \frac{\eta_R I_1(r/\delta)}{\eta I_2(R/\delta)+\eta_R I_0(R/\delta)}2\omega_p\delta, &\textrm{convex}\\
    -\frac{\eta_R K_1(r/\delta)}{\eta K_2(R/\delta)+\eta_R K_0(R/\delta)}2\omega_p\delta, & \textrm{concave} \label{appeq:tangential_velocity_curv}
    \end{cases}
\end{equation}

Note, that the modified Bessel function also forms an exponential decay. If the cluster radius is a few times larger than the decay length of the edge flow, the Bessel function is very close to the exponential decay, as we show in Fig.~\ref{fig:Bessel_vs_exponential}. All clusters and voids in our experiments have a radius to decay length ratio $R/\delta\gtrsim5$, meaning they are large enough to be modeled by the exponential function in Eq.~(2) in the main text.

\section{Derivation of bulk cluster rotation}

\label{app:bulk_rotation}

In this Appendix, we derive the bulk angular velocity of a cluster induced by shear stress from the edge flows. We assume that only a thin outer layer of particles moves with velocity $v_\varphi(R)$. The friction between this layer and the bulk of the cluster imposes a torque

\begin{equation}
    \tau_1=2\pi R \cdot v_\varphi(R)\eta \cdot R,
\end{equation}
where $2\pi R$ is the area of the interface between the outer layer and the bulk and $v_\varphi(R)\eta$ is the stress applied by the outer layer onto the cluster.
Additionally, the cluster exhibits a friction with the substrate which induces a torque
\begin{equation}
    \tau_2 = \int\limits_0^R dr 2\pi r \cdot \Gamma_v\Omega r \cdot r=\Omega \Gamma_v\pi R^4 / 2,
\end{equation}
where at each layer with radius $r$ and width $dr$ the area of the contact with the substrate is $2\pi rdr$ and the stress is $\Gamma_v\Omega r$, with $\Omega$ being the angular velocity of the cluster.

We are looking for $\Omega$ which balances the two torques
\begin{equation}
    \Omega = \frac{4v_\varphi(R)\eta}{\Gamma_v R^2}\propto R^{-2} v_\varphi(R). \label{appeq:bulk_rotation}
\end{equation}
When edge velocity weakly depends on the cluster radius, as is the case in our system, bulk rotation decays with cluster radius as $R^{-2}$. We demonstrate in Fig.~\ref{fig:bulk_rotation_vs_radius} that the fitted angular velocities of our clusters follow this prediction.\\

\section{Data evaluation and fitting procedure}

\label{app:data_evaluation}

Particle tracking was accomplished via IMARIS software. Using the Brownian tracking algorithm, individual particle positions and average speeds were acquired. The average speed is the total traveled {path} over the time of the video. Positions were computed relative to the edge of the cluster/void and were averaged with time as well. The size of the clusters/voids was also calculated through the IMARIS software taking advantage of the surface function. An effective radii for each cluster/void was estimated from a measured 2-D surface area.

Positions relative to the edge, computed with IMARIS software, had values larger than the radius of a single particle. In order to obtain distances from the edge that are zero for particles closest to the edge, the distance data was corrected. For this, a minimal distance over all particles was subtracted from the distance of each particle $x_i\to x_i - \min\limits_i x_i$.

\begin{figure}
    \centering
    \includegraphics{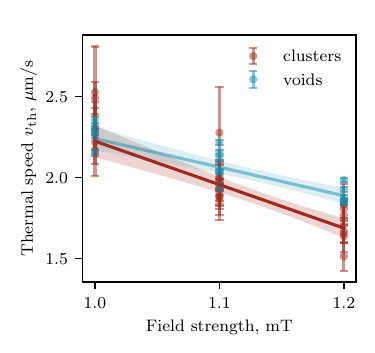}
    \caption{Thermal speeds $v_\mathrm{th}$ obtained from speed fits using Eq.~2 and 4 from the main text for voids (blue dots) and clusters (red dots), respectively. The solid lines indicate linear fits with 95\% confidence intervals (shaded).}
    \label{fig:vth_field}
\end{figure}

The average speeds and positions of particles in clusters/voids were fitted to an exponentially decaying function with an offset
\begin{equation}
\label{appeq:model}
    v(x) = v_\textnormal{max}e^{-x/\delta} + v_\textnormal{th},
\end{equation}
with a maximal edge {speed} $v_\mathrm{max}$, decay length $\delta$, and thermal {speed} $v_\mathrm{th}$ being the fitting parameters.
To reduce the number of fitting parameters, we followed a two step procedure. (i) We fit the data with Eq.~\eqref{appeq:model} for several clusters/voids of different size at a fixed field strength. We then compute an average decay length $\delta$ and thermal {speed} $v_\mathrm{th}$ over these clusters/voids. As decay length and thermal {speed} are assumed to be independent of system size, their average values were taken as true estimates. (ii) Fixing the decay length and thermal {speed} to their average values, we fit the maximal edge {speed} $v_\mathrm{max}$ again using Eq.~\eqref{appeq:model}. The resulting parameters are shown for different clusters and voids at three different field strengths in Figs.~2, 3 and 4 in the main text.

In clusters, the fit was additionally corrected by taking into account the bulk rotation:
\begin{equation}
    v(x)=v_\textrm{max}e^{-x/\delta}+v_\textnormal{th}+\Omega(R-x).
    \label{appeq:rotfit}
\end{equation}
Here, $\Omega$ is the bulk cluster rotation and $R$ is the effective radius of the cluster determined by fitting the cluster area to a representative circle. The fitting procedure was again split in two steps. (i) We fit the data with Eq.~\eqref{appeq:rotfit} for several clusters of different size at a  fixed field strength. We then compute an average decay length $\delta$ and thermal {speed} $v_\mathrm{th}$. (ii) Fixing these parameters to the average values, we fit the maximal edge {speed} $v_\mathrm{max}$ and the bulk cluster rotation $\Omega$ again using Eq.~\eqref{appeq:rotfit}.

To confirm that the bulk rotation term is necessary to better describe particle {speeds} in clusters, we compare two fits, with and without bulk rotation, for a representative cluster. We choose a cluster at an intermediate value of magnetic field $H_0=11$  mT and fit Eqs.~\eqref{appeq:model} and \eqref{appeq:rotfit} to its mean {speeds} in Fig~\ref{fig:cluster_two_fit}. We see that residuals $v_\mathrm{data}-v_\mathrm{fit}$ are more symmetric around zero, and the sum of their squares is smaller in the fit with bulk rotation than in the fit without rotation. This supports our claim in the main text that the bulk rotation term is necessary to describe macroscopic dynamics in clusters.

\begin{figure}
    \centering
    \includegraphics{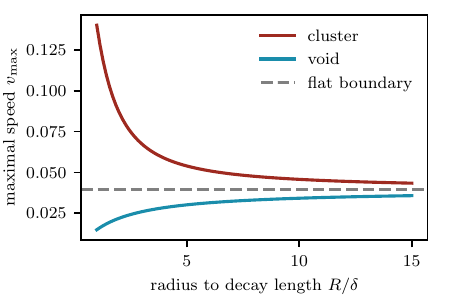}
    \caption{Maximal speed on the edge of a cluster (red) or void (blue) as a function of its radius, predicted from the modified Bessel function solution of the hydrodynamic equation. Parameters are fixed to be $\omega_p=1$, $\delta=1$, $\eta/\eta_R=50$. Dashed line indicates the maximal speed at flat boundary which is achieved at $R\to\infty$.}
    \label{fig:vmax_vs_radius}
\end{figure}

\begin{figure*}
    \centering
    \includegraphics[scale=0.95]{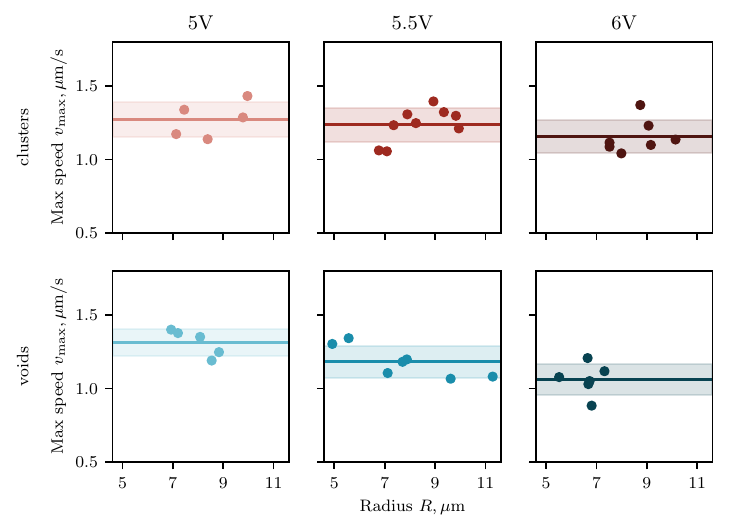}
    \caption{Plot fitted maximal speed in clusters and voids as a function of cluster or void radius for three different magnetic field strengths. Solid lines show averages over all clusters or voids of different radii. Shaded regions show the corresponding standard deviation.}
    \label{fig:vmax_fit_vs_radius}
\end{figure*}

\begin{figure*}[h]
    \centering
    \includegraphics[scale=0.95]{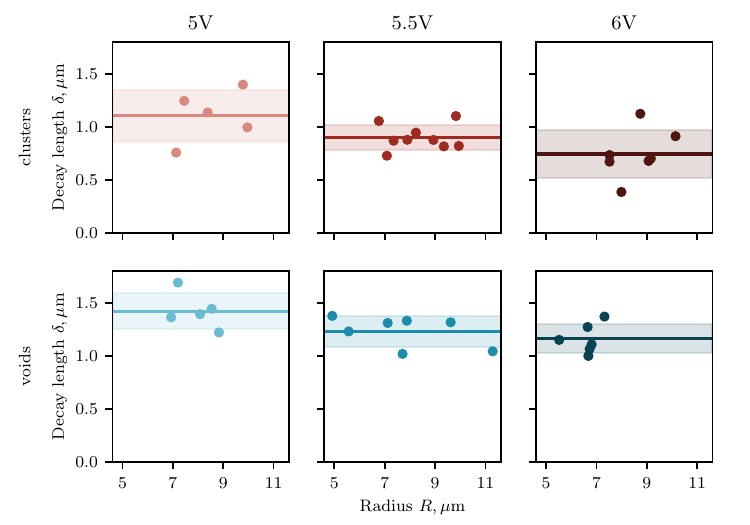}
    \caption{Plot fitted decay length in clusters and voids as a function of cluster or void radius for three different magnetic field strengths. Solid lines show averages over all clusters or voids of different radii. Shaded regions show the corresponding standard deviation.}
    \label{fig:declen_fit_vs_radius}
\end{figure*}

\section{Thermal speed decays with field strength}
\label{app:thermal_speed}

In this Appendix, we demonstrate that thermal speed in both clusters and voids decays with the field strength, as visible in Fig.~\ref{fig:vth_field}.

\section{Maximal edge speed and decay length are independent of cluster or void radius}

\label{app:max_edge_speed}

In this Appendix, we demonstrate that the maximal edge speed and the decay length of edge flows do not depend on the system size for our clusters and sheets with voids.

First, we discuss the case of the maximal edge speed. We consider the solution of the hydrodynamic equations with curved boundary given by Eq.~\eqref{appeq:tangential_velocity_curv}. We plot the maximal velocity in this solution as a function of cluster and void radius in Fig.~\ref{fig:vmax_vs_radius} for all other parameters being fixed. The edge velocity saturates at a constant value in large systems. This value corresponds to a system with flat surface. In the studied systems, the radius is large enough such that the dependence of maximal edge velocity on radius is negligible (Fig.~\ref{fig:vmax_fit_vs_radius}). This allows us to treat all clusters and voids as if they have a flat surface.

We further show that the decay length $\delta$
does not depend on the size of clusters or voids, as we illustrate in Fig.~\ref{fig:declen_fit_vs_radius}. The difference in decay length between clusters and voids, observed in Fig.~4(a) of the main text, happens not due to the difference in their sizes, but because voids are constrained, while the clusters are not.

\bibliography{references_sync}{}

\end{document}